\documentclass[aps,prf,superscriptaddress,preprint]{revtex4-2}

\usepackage[T1]{fontenc}
\usepackage[latin9]{inputenc}
\usepackage{graphicx}
\usepackage{babel}

\usepackage{epsfig,amssymb,amsmath,graphicx,subfigure,hyperref}  
\usepackage{color}
\usepackage{multirow}
\usepackage{tabularx}
\usepackage{mathrsfs}
\usepackage{bm}
\usepackage{float}

\begin{document}

\title{Robust propagation of internal coastal Kelvin waves in complex domains}


\author{Chenyang Ren}
\affiliation{School of Physics and Astronomy and Institute of Natural Sciences, Shanghai Jiao Tong University, Shanghai 200240, China}
\author{Xianping Fan}
\affiliation{School of Physics Science and Engineering, Tongji University, Shanghai 200092, China}
\author{Yiling Xia}
\affiliation{Ocean University of China, Qingdao 266100, China}
\author{Tiancheng Chen}
\affiliation{School of Physics Science and Engineering, Tongji University, Shanghai 200092, China}
\author{Liu Yang}
\affiliation{School of Physics and Astronomy, University of Manchester, Manchester M139PL, United Kingdom}
\author{Jin-Qiang Zhong}
\email{jinqiang@tongji.edu.cn}
\affiliation{School of Physics Science and Engineering, Tongji University, Shanghai 200092, China}
\author{H. P. Zhang}
\email{hepeng\_zhang@sjtu.edu.cn}
\affiliation{School of Physics and Astronomy and Institute of Natural Sciences, Shanghai Jiao Tong University, Shanghai 200240, China}
\affiliation{Collaborative Innovation Center of Advanced Microstructures, Nanjing,
China}



\date{\today}

\begin{abstract}
We experimentally investigate internal coastal Kelvin waves in a two-layer fluid system on a rotating table. Waves in our system propagate in the prograde direction and are exponentially localized near the boundary. Our experiments verify the theoretical dispersion relation of the wave and show that the wave amplitude decays exponentially along the propagation direction. We further demonstrate that the waves can robustly propagate along boundaries of complex geometries without being scattered and that adding obstacles to the wave propagation path does not cause additional attenuation. 
\end{abstract}


\maketitle

Over recent decades, tools from topology have shed new light on a wide range of physical phenomena. One of such phenomena is the emergence of robust boundary states under topological protection. For example, Hall current in two-dimensional semiconductors can robustly and unidirectionally propagate along the boundaries which may have defects and complex shapes; its persistence is guaranteed by topological arguments \citep{kane2005z,kane2005,hasan2010,bernevig2006,halperin1982}. Topologically protected states have also been found in classical systems, including two-dimensional chiral materials \citep{J.C.Tsai2005,Nash2015,soni2019,Xiang2020}, isostatic lattices in two dimensions \citep{kane2014}, and photonic systems \citep{khanikaev2017}.

Recently, fluid flow phenomena have been examined with topological methods \citep{Faure,delplace2020,Delplace2017,tauber2019,perrot2019,Tauber2020}; examples include oceanic Kelvin waves \citep{Delplace2017,tauber2019} and atmospheric Lamb waves \citep{perrot2019}. Robust boundary states, behaving like the topologically protected states, have also been found in highly nonlinear systems, such as the wall states in the rotating Rayleigh--B{\'e}nard system \citep{favier2020}. Analysis of the shallow-water model \citep{Delplace2017} shows that the earth's rotation breaks the time-reversal symmetry and leads to a gap between the low and high frequency wave bands. Eastward-propagating Kelvin waves with frequencies in the gap can be considered as topologically protected boundary states at the equator, where the earth's rotation changes signs \citep{Delplace2017}. Coastal Kelvin waves trapped near a sharp boundary have also been analyzed from a topological point of view \citep{Tauber2020}, because the robust propagation of these waves along irregular coasts has been well documented \citep{Mysak2006,pedagogical}. Current theoretical analysis has not been able to rigorously establish coastal Kelvin waves as topologically protected boundary states and topological origin of these boundary waves remains to be explored \citep{Tauber2020}. 

As an important geophysical phenomenon, coastal Kelvin waves have been extensively studied by theoreticians to explore the effects of boundary conditions, stratification, and bottom topography \citep{iga1995,allen1975,kawabe1982,killworth1989}. In addition, experimental studies of coastal Kelvin waves, which have some technical challenges, have also been carried out to investigate modal wave structures, evolution of nonlinear waves, and reflection of solitary waves \citep{caldwell1972,Ulloa2014,Renouard1995}. In this work, we focus on linear wave propagation phenomenon and possible interactions between linear waves and obstacles. To that end, we set up a two-layer fluid system on a rotating table and use a computer-controlled wave-maker to excite coastal Kelvin waves. Wave propagation along the fluid interface is quantified by particle image velocimetry (PIV). Our measurements show that the wave in our system propagates in the prograde direction while decaying exponentially along the path and that measured dispersion relation is in good agreement with inviscid theory prediction for low-frequency waves. We further demonstrate that internal coastal Kelvin waves robustly propagate along various irregular boundaries without being scattered into the bulk and that the complex geometry of the boundaries doesn't lead to additional wave dissipation. 

\begin{figure}
	\centering
	\includegraphics[width=0.8\textwidth]{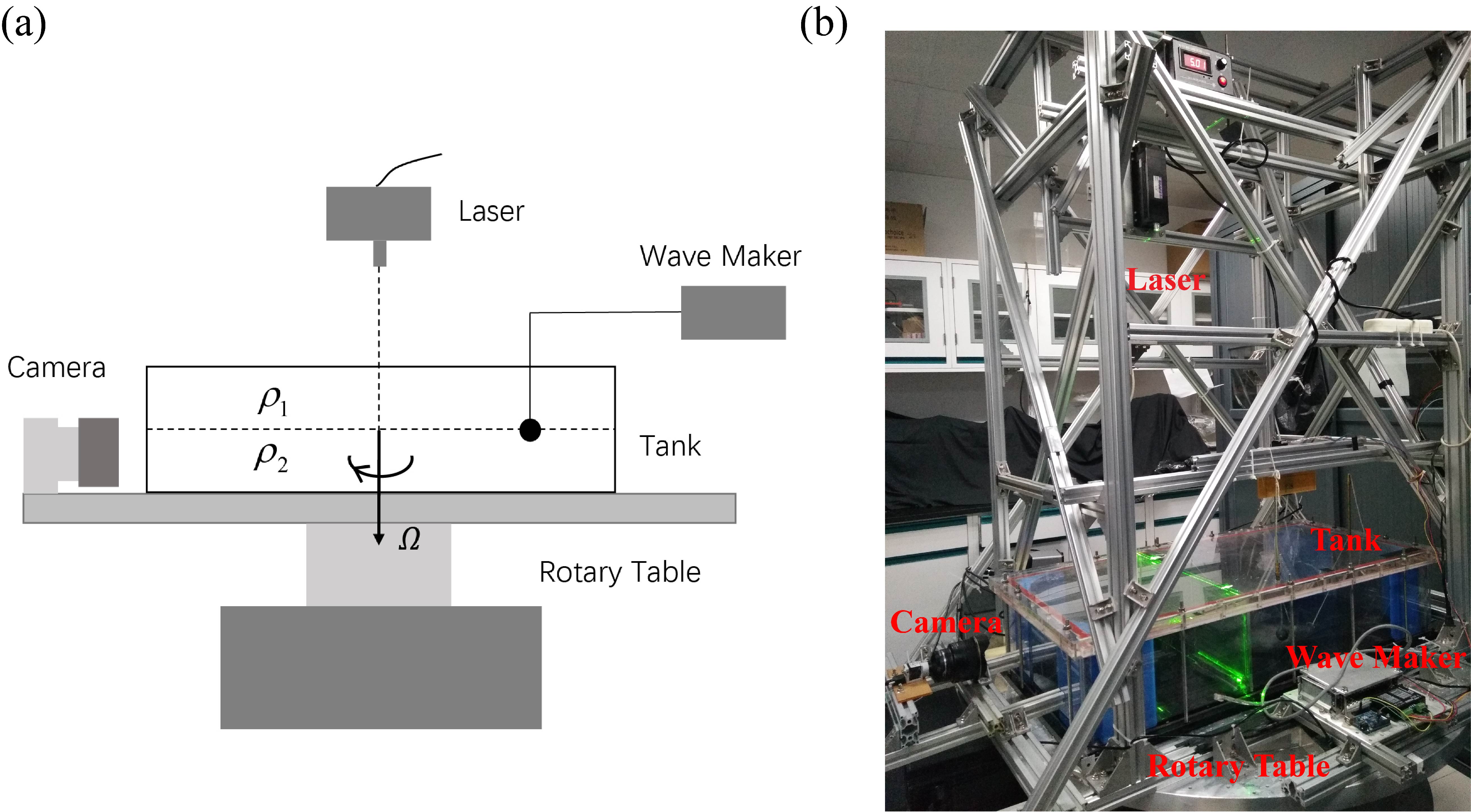}
	\caption{Schematic diagram and photography of the experimental setup. The various components are explained in the text.}
\end{figure}

A schematic drawing and a photograph of the experimental apparatus are shown in Fig. 1. It consists of a rotary table, a fluid tank, a wave-maker and a Particle Image Velocimetry (PIV) system for flow-field measurements. The rotary table is fixed securely on the floor. Both its rotating rate $\Omega$ (clockwise) and rotational acceleration $a$ can be adjusted precisely.
We use a rectangular fluid tank that has inner dimensions of $1000\times500\times200$ (mm$^3$). Its bottom plate, made of oxidized aluminum, is mounted on the rotary table. Its sidewalls, as well as the top lid, are made of 10 mm thick transparent Plexiglas plates. The tank has four rounded corners in a radius of 150 mm, designed to eliminate the corner vortices during the spin-up of the fluid and reduce fluid mixing. The tank is filled with two fluid layers, both are 100 mm in thickness. The upper fluid layer is fresh water and the lower one is saltwater. Their densities are $\rho_1{=}0.998$ g/cm$^{3}$ and $\rho_2{=}1.020$ g/cm$^{3}$, respectively. In order to prepare the two fluid layers with sharp density interface, we first fill the lower half of the tank with the salty water. A low-density sponge in size of $150\times150\times15$ (mm$^3$)  is gently placed on the salty water surface. Fresh water at a very slow flow rate of $3.09\times10^{-4}$ m/s is injected on top of the sponge. It permeates through the sponge, and then flows steadily onto the fluid tank. During this filling process \citep{Whitehead2007}, we observe no noticeable mixing between the top and bottom fluid layer, and two fluid layers with an interfacial thickness $\delta\approx2$ mm is prepared. After the tank is filled, the rotary table spins up to a rotation rate $\Omega$ with a steady acceleration $a$. For all experiments we choose $\Omega$=1.88 rad/s and $a{=}2.62\times10^{-4}$ rad/s$^2$. The acceleration $a$ is low enough to prevent mixing between the two fluid layers. The fluid system then reaches a state of solid-body rotation after the time-scale of spin-up, $\tau{=}L/{(}\nu\Omega{)}^{1/2}$, which is approximately 20 minutes in the present experiment. Here $L$ is the long dimension of the tank and $\nu$ is the fluid viscosity. 

The internal waves are excited by a wave-maker made of a rigid sphere. The sphere has a radius $R{=}15$ mm, connected through an electric guide rail to a step motor. During the experiment, the sphere is driven to oscillate vertically in the form of a cosine function $h(t){=}h_0+H$cos$(\omega t)$. The equilibrium position $h_0$ of the sphere is selected so that the sphere center locates right at the interface of the two layers. For all measurements, the center of the sphere is 18 mm away from the long lateral wall of tank. We use an Arduino micro-controller to control the oscillation frequency $\omega$ and amplitude $H$ of the wave-maker. 

\begin{figure}
	\centering 
	\includegraphics[width=0.8\textwidth]{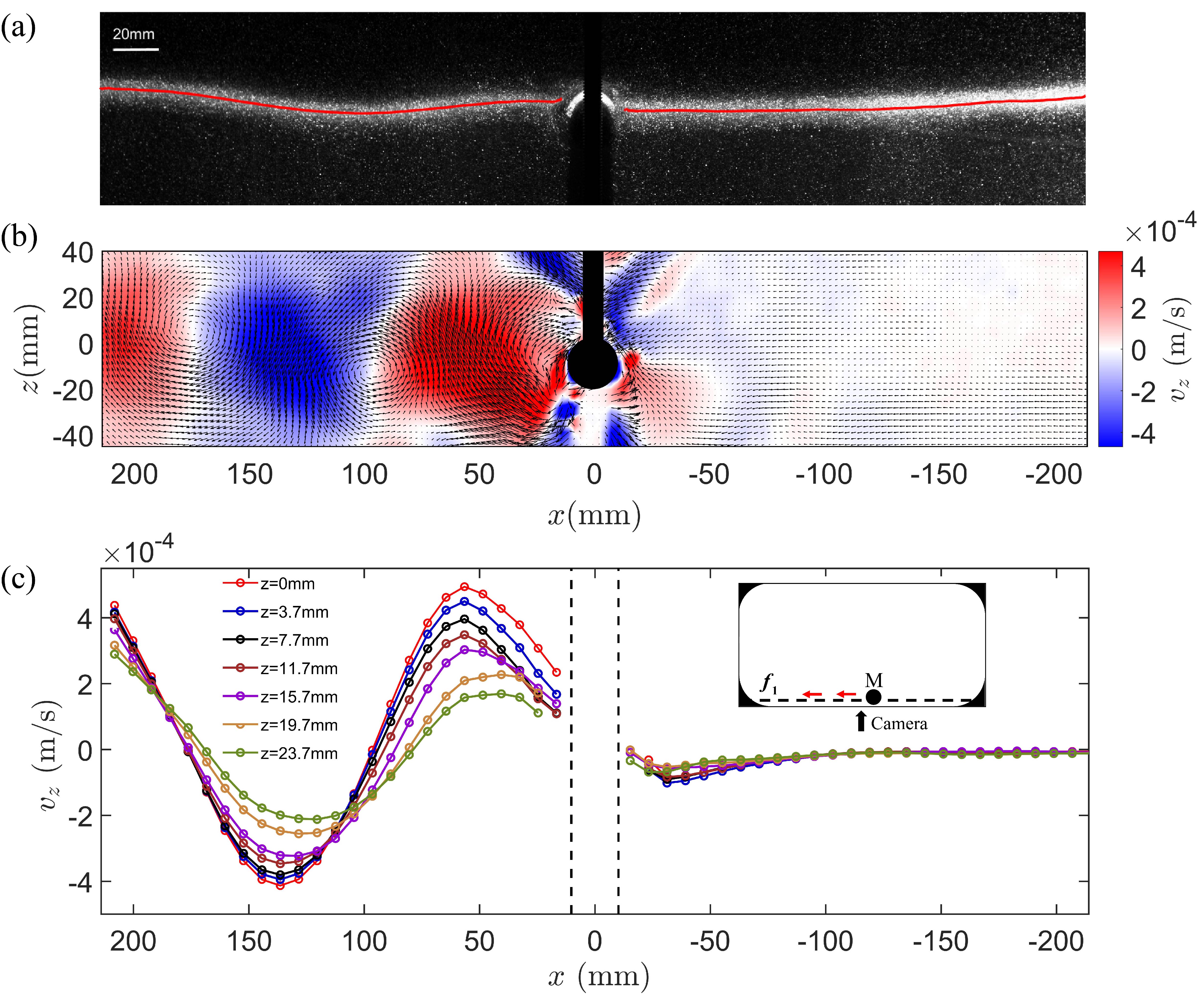}
	\caption{
		(a)	A snapshot of seeding particles in a vertical plane $f_1$ (see inset). (b) The corresponding instantaneous velocity field extracted from our PIV system. The background coloration corresponds to the vertical velocity component $v_z(x,z)$. The central dark region denotes the position of the wave-maker. We define the positive horizontal direction of $x$ to be leftward, i.e., the propagation direction of the waves. (c) The vertical velocity component $v_z(x)$ measured at various fluid heights. Inset: a schematic drawing of top view of tank. The circle M denotes the position of wave-maker. The dashed line $f_1$ is the position of the laser light-sheet in PIV measurements. The view angle of camera is indicated by black arrow. The red arrows present the propagation path of waves. 
}
\end{figure}

We take velocity measurements using a PIV system installed on the rotary table \citep{shi2020fine}. A thin vertical light-sheet (1 mm in thickness) powered by a continuous solid-state laser illuminates the seeding particles over various locations in the fluid tank (insets in Figs. 2-5). Hollow glass beads, with a diameter of 50 $\mu$m and an effective density of 1.030 g/cm$^3$, are used as the seeding particles. Since these particles sediment three times faster in fresh water than in salty water, they accumulate at the interface during the spin-up process and form a bright region in raw particle images. From the bright region, we can track the fluid interface $\eta(x)$, see red line in Fig. 2(a). Images of the particles are captured at a frame rate of 10 fps by a high-resolution Basler acA2040-90um camera ($2048\times2048$ pixels) mounted in the co-rotating frame. Two-dimensional velocity maps are obtained by cross-correlating two consecutive images taken at a time interval according to the flow speeds. Each velocity vector is calculated from an interrogation windows ($32\times32$ pixels), with $50\% $ overlap of neighboring sub-windows \citep{Westerweel2013}. We choose the measurement area as a rectangular region of $433\times85$ (mm$^2$) crossing the interface of the two fluid layer, reaching a spatial resolution of 1.33 mm in the velocity fields.

We first examine the property of unidirectional propagation of the Kelvin wave \citep{wang2002}. For this purpose, we install the wave-maker at the middle of the long side of the tank (position M in inset of Fig. 2(c)). The fluid velocity over a vertical plane $f_1$ is measured. Figure 2(a) shows an example of the particle density field captured through our PIV system with the oscillation amplitude $H$=20 mm and frequency $\omega{=}1.90$ rad/s. The interface of the two fluid layers $\eta(x)$ is marked by the red curve, that is determined by the average height of the bright area for each horizontal position $x$. It shows oscillations on the left side of the wave-maker, but a parabolic profile $\eta(x){=}\eta_0(x){=}bx^2$ on the right, caused by the centrifugal force (See Supplemental Material \citep{supplementary}).  We choose the vertical interfacial position at $x{=}0$ as the zero of the fluid height, i.e., $\eta_0(x{=}0){=}0$.  These are direct observations that the internal wave indeed propagates leftward unidirectionally. (See Supplementary Movie for the propagation of the Kelvin wave).

Figure 2(b) presents the instantaneous velocity field $v(x, z)$ extracted from the PIV system. A prominent feature shown in the velocity field is the apparent periodic oscillations of both the velocity components $v_x$ and $v_z$ on the left for $x{>}R$. The magnitude of fluid velocity on the right for $x{<}-R$, however, is nearly zero. In the region of $|x|\le R$, the fluid flow is complex due to the large-amplitude disturbances of the oscillation source. Along the interface $z{=}\eta(x)$ for $x{>}R$, we find that the maxima and nodes of $v_z$ appear alternatively, while $v_x$ remains close to zero. In Fig. 2(c) we plot the vertical fluid velocities $v_z$ against the horizontal position $x$ for various fluid heights. Velocity data for the oscillation-source region ($|x|\le R$) are excluded. For $x{>}R$ the velocity profiles $v_z(x)$ can be best described by an oscillation function with its amplitude decaying exponentially $v_z(x){=}v_z^0(z)$exp$({-}\alpha_v x)$cos$(kx{-}\omega t)$. Here $v_z^0$ is the oscillation amplitude that depends on $z$ and is maximum at $z$=0. $k$ is the wave number and  $\lambda_{v_d}{=}\alpha_v^{-1}$ is the decay length of $v_z$. For each fluid height $z$ the crest and trough of $v_z$ appears periodically. For $x{<}-R$, no significant velocity component is observed ($v_z\approx 0$). We stress that such a dynamical feature of unidirectional propagation has been observed in a wide range of the experimental parameters ($\omega$, $H$). 

\begin{figure}
	\centering
	\includegraphics[width=1\textwidth]{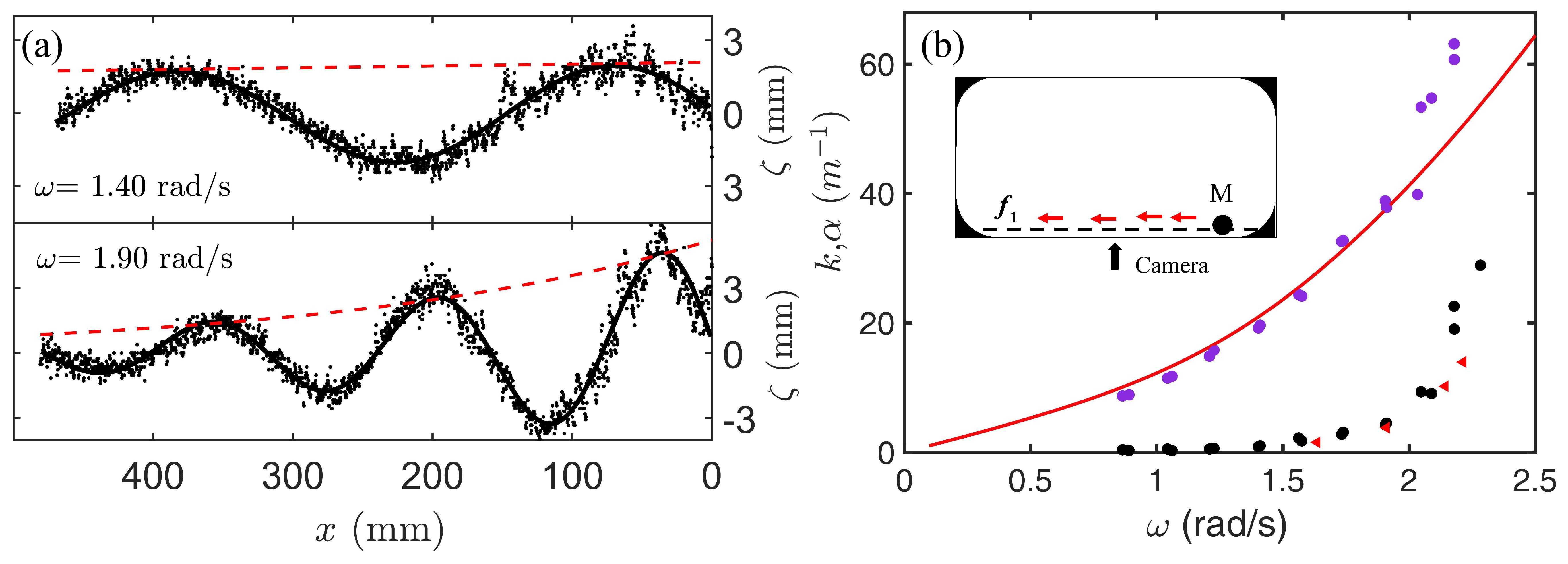}
	\caption{(a) The interfacial profiles $\zeta(x)$ of Kelvin waves propagating along a straight fluid boundary. Results for $H$=15 mm, $\omega$=1.40 rad/s, and 1.90 rad/s. Black dots: the raw data of $\zeta(x)$. Black solid lines: fitted curves of $\zeta(x)$ according to formula (1). Red solid lines: the amplitude $A$ as a function of $x$, $A=A_0$exp$(-\alpha x)$. (b) The measured damping cofficient $\alpha$ (black dots for $H$=15 mm and red triangles for $H$=20 mm) and wave number $k$ (purple) as functions of the oscillation frequency $\omega$. Red solid line: dispersion relation expressed by formula (2). Inset: a schematic drawing of top view of tank.
}
\end{figure}

 To further illustrate the propagation dynamics of internal coastal Kelvin wave, we move the wave-maker to the right side of tank (position M in inset of Fig. 3(b)), leaving a large distance for the leftward propagation of the Kelvin waves. We perform a set of measurements of the interfacial position $\zeta(x)$,  choosing the range of the wave frequency as 1.30 rad/s $\le\omega \le$ 2.40 rad/s. As shown below, wave in this frequency range can be conveniently measured in our tank because of their wavelength and decaying length. 

Figure 3(a) shows the relative interfacial positions $\zeta(x){=}\eta(x){-}\eta_{0}(x)$ for various frequencies. 
Here $\eta_0(x){=}bx^2$ is the parabolic interface measured when the wave-maker is turned off. Our experimental data of $\zeta(x)$, which are shown in the black dots, agree well with the following function:
\begin{align}
\zeta(x,t)&=A_0\exp(-\alpha x)\text{cos}(kx-\omega{t}+\phi), \label{eq1}
\end{align}
as shown by the fitted black curves. $\alpha$ is the damping cofficient and $k$ is wavenumber. The wave amplitude decays exponentially along the prorogation direction $x$ as implied by the red lines; for progressive linear interfacial waves, dissipation mainly occurs at the sidewalls and interfacial boundary layer \citep{Troy2006}. Figure 3(b) shows the experimental data of $k$, $\alpha$ as functions of $\omega$. With increasing $\omega$, both $\alpha$ and $k$ increase. In the range of $\omega\le1.90$ rad/s, we find that the wavenumber $k$ is about one order in magnitude larger than the damping coefficient  $\alpha$ \citep{Troy2006,mitsudera1989,smirnov2011,Pinsent2006}, meaning that viscous effects are weak for these waves. The dispersion relation for coastal Kelvin wave can be solved in a inviscid two-layer model \citep{1979Non} (also see supplementary materials \citep{supplementary}):

\begin{align}
 \omega^2&=\frac{gk(\rho_2-\rho_1)}{\rho_1\coth{(kD_1)}+\rho_2\coth{(kD_2)}},\label{eq2}
\end{align}
where $D_1, D_2$ is the depth of two-layer fluids respectively. The red line in Fig. 3(b) shows the theoretical results, without free parameters, from Eq. (\ref{eq2}), which is in close agreement with the experimental data in the low-frequency range when  $\omega\le1.90$ rad/s. Therefore, the inviscid dispersion relation is confirmed in low-frequency waves. We also explore the influence of forcing amplitude $H$ on $\alpha$; as shown in Fig. 3(b), the $\alpha$ parameters measured with two amplitudes are nearly identical. The observed amplitude-independence of the damping coefficient suggests that the Kelvin waves in our experiments propagate in the linear regime, which is consistent with the dispersion relation for linear waves measured in Fig. 3(b). We note that nonlinear phenomena, such as solitary waves and triads interactions, do not occur in our experiments in the linear regime.

\begin{figure}
	\includegraphics[width=0.7\textwidth]{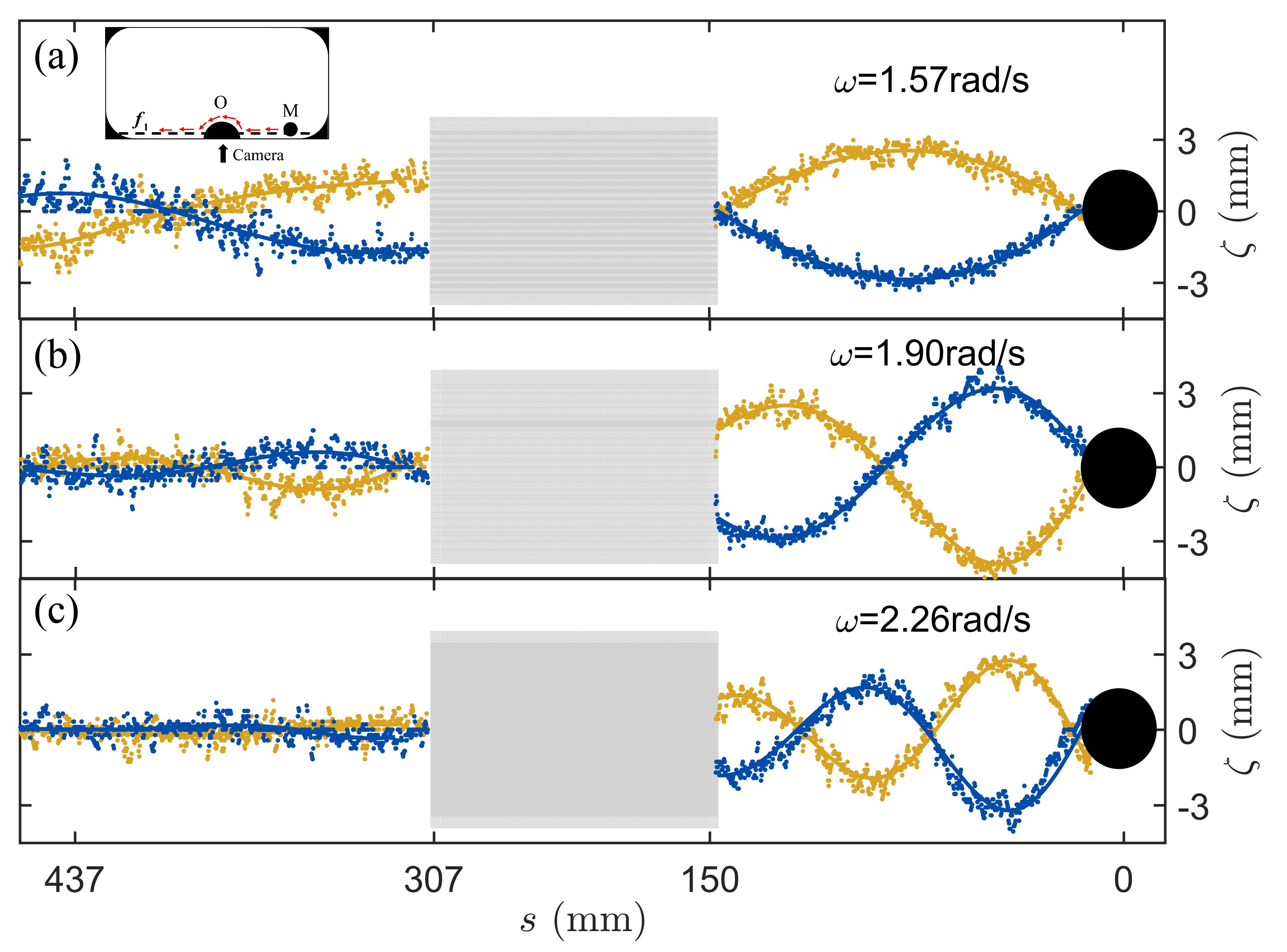}
	\centering
	\caption{
		The interfacial profiles $\zeta(s)$ of Kelvin waves in plane $f_1$ when it propagates surrounding an semi-cylindrical obstacle. The experiment is performed with $H$=15 mm, $\omega $=1.57 rad/s (a), 1.90 rad/s (b) and 2.26 rad/s (c), respectively. Dots: the raw data of $\zeta(s)$. Solid lines: fitted curves of $\zeta (s)$ using formula (1). The $s$-coordinate presents the travel distance of the wave. Black circle: position of the wave-maker. Gray rectangular: position of the semi-cylindrical obstacle. Inset:  a schematic drawing of top view of tank. The diameter of the semi-cylindrical obstacle O is $100$ mm.  
	}
\end{figure}

For high frequencies $\omega{>}1.90$ rad/s, the damping coefficient $\alpha$ increases sharply with increasing $\omega$. In this case, the  theoretically predicted dispersion relation is not accurate because of the strong frictional damping, and  the wavelength $\lambda_L=2\pi/k$ and decay length $\lambda_d{=}\alpha^{-1}$ become comparable. This is shown in Supplementary Movie S2 \citep{supplementary}: high-frequency wave decays rapidly along the propagation path. Hence, in our following experiments, we mainly use the wave with frequency $\omega\le1.90$ rad/s  to further investigate  the robustness of the coastal Kelvin wave propagation in complex fluid domains \citep{Tauber2020}.

In our experimental settings, we have installed obstacles (approximately 5 times larger than the Rossby radius of deformation in size) of different geometric shapes in the fluid tank along the propagation path of the Kelvin wave. These obstacles change the geometry of the fluid boundary as well as the propagation path, but do not alter the level of topological complexity of the system \citep{kane2005z,kane2005,hasan2010,bernevig2006,halperin1982}. 

\begin{figure}
	\centering 
	\includegraphics[width=1\textwidth]{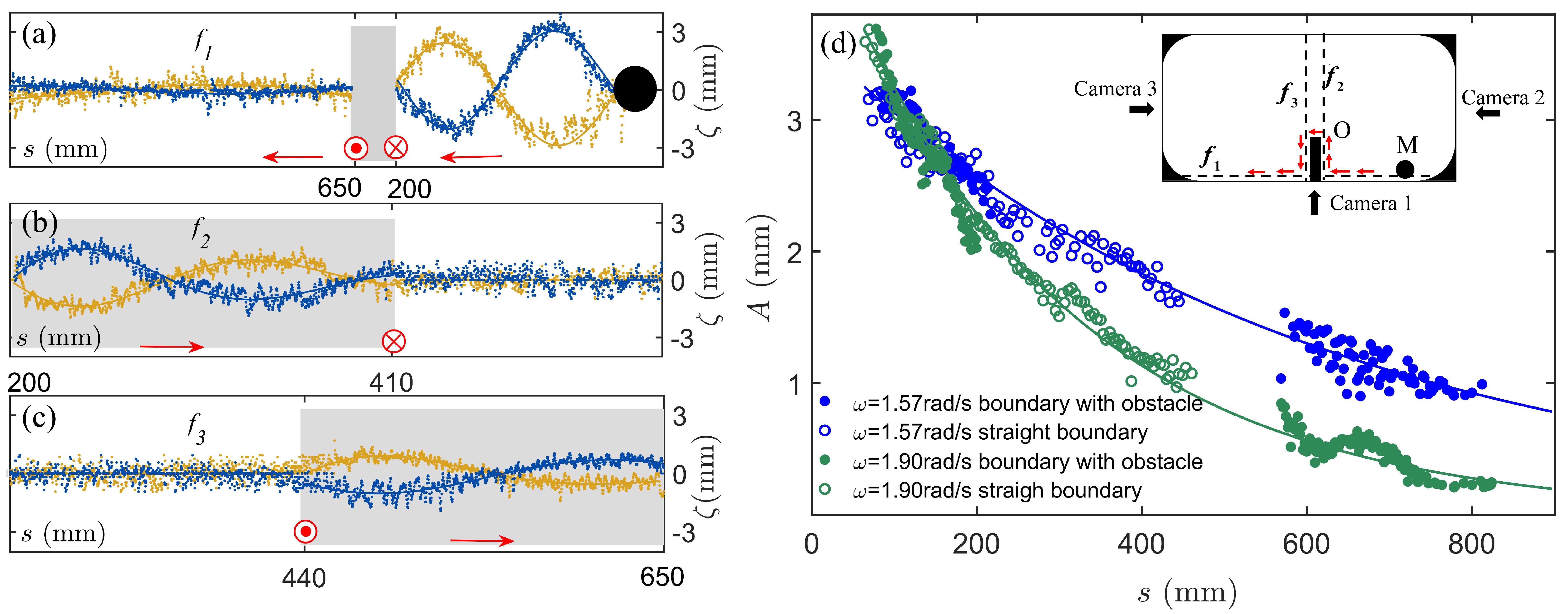}
	\caption
	{  (a-c) The interfacial profiles $\zeta(s)$ of Kelvin waves propagating surrounding a thin rectangular obstacle (see inset). A $30$ mm thick obstacle board O with a length $L_c$=210 mm is installed ,  perpendicular to the lateral wall of the tank (inset). Dots: the raw data of $\zeta(s)$. Solid lines: fitted curves of $\zeta(s)$ using formula (1). The red arrows and circles indicate the direction of wave propagation. Black circle: position of the wave-maker. Gray rectangular: position of the obstacle board.
		(d)	The wave oscillation amplitudes $A(s)$ as a function of the travel distance $s$. Symbols: data measured with an obstacle board in length $L_c$=110 mm. The solid lines present exponential function $A(s){=}A_0$ exp($-\alpha s)$ fitted to the straight-boundary data (open symbols). Inset: a schematic drawing of top view of tank. The dashed lines $f_1$, $f_2$, $f_3$ are the positions of the laser light-sheet in PIV measurements, captured by the camera with corresponding number at different angles (black arrows).
	}	 
\end{figure}

In a series of measurements, we first install a semi-cylindrical obstacle O near the wall, as shown in the inset of Fig. 4(a). The relative interfacial position $\zeta$ for the waves of three frequencies $\omega$ is measured in the vertical plane $f_1$ at two times separated by half an oscillation period. We see that with low frequencies $\omega$=1.57 rad/s and $\omega$=1.90 rad/s, the Kelvin wave can travel leftward around the obstacle and reach the wall to the left of obstacle, rather than being reflected by the obstacle. The Kelvin wave with a high frequency, $\omega$=2.26 rad/s, however, decays to almost zero, when it arrived to the left of obstacle, because of the strong energy dissipation under large frequency, as discussed in Fig. 3. 

For further demonstration of the robustness of near-shore propagation of Kelvin wave, we install a barrier board O that is right-angle to the long lateral wall of the tank (see inset of Fig. 5(d)). The fluid interface is measured along all sections of the wave propagation paths, i.e., in the vertical planes of $f_1$, $f_2$ and $f_3$. In Fig. 5(a)-(c) we present the interfacial positions $\zeta(s)$ captured at two oscillation phases that are half-period apart in planes $f_1$, $f_2$ and $f_3$ respectively. The experiment is performed with $\omega$=1.73 rad/s, $H$=15 mm. Figure 5(a) shows within a short travel distance $s{<}200$ mm that the energetic Kelvin wave is propagating leftward. As the wave comes across the barrier board O at $s$=200 mm, figures 5(a) and 5(b) indicate that rather than being reflected backward, the wave makes a sharp right turn of 90 degrees and then travels in the direction perpendicular to the long lateral wall of the tank along board O. Another striking behavior is observed when the wave arrives at the end of board O at 410 mm$\le s \le$ 440 mm. Figures 5(b) and 5(c) illustrate that at this location the wave changes again sharply its propagation direction and turns around the edge of board O clockwisely, with its propagation path nested inside the vicinal region next to the obstacle surface. This robust near-shore flow  continuously returns to the tank wall at $s$=650 mm and then propagates leftward, although its amplitude has significantly decayed shown in Fig. 5(a).  This robust and unidirectional propagation of the coastal Kelvin wave shows that its dynamics is insensitive to the details of the boundary geometry \citep{delplace2020,Tauber2020}. In addition, we measure the profile of internal Kelvin waves in the direction perpendicular to the boundary (see supplementary materials for Figure S1 \citep{supplementary}). Our measurements show that the internal Kelvin wave is trapped exponentially near the boundary \citep{wang2002}.
For a wave of frequency $\omega=1.57$ rad/s, Fig. S1 shows a decaying length (in the y-direction) of $19.56\pm2.02$ mm. This is consistent with a direct estimation of the Rossby radius of deformation $\Lambda={\omega}/{2k\Omega}$ with the wave frequency and vector data from Fig. 3(b) \citep{Kundu2002Fluid}.

The general topological property of the Kelvin waves is further investigated by measuring the wave amplitudes along various traveling paths surrounding different obstacles. As pointed out in Fig. 3, the amplitude of the Kelvin wave decays exponentially with increasing travel distance due to the viscous damping. Here, we examine whether the decay length $\lambda_d$ is independent of the path of wave propagation. In Fig. 5(d) we show the wave amplitude $A$ as a function of the propagation distance $s$ measured under various settings of boundary conditions and different oscillation frequencies $\omega$. The open symbols represent the results for Kelvin waves propagating along a straight lateral wall, while the filled symbols are the results obtained when the Kelvin waves bypass along obstacles. We find that the decay length $\lambda_d$ decreases with increasing $\omega$. For a fixed $\omega$, the two sets of data collapse approximately into a single curve $A(s){=}A_0$exp$(-\alpha s)$. These results imply that the decrease of the wave amplitude $A$ along a propagation path is determined by the traveling distance $s$, but independent on the geometry of the path. Previous theoretical studies predicted that the Kelvin waves with frequencies of $\omega$<$2\Omega$ have no additional attenuation of amplitude in the process of diffraction at different barriers, by means of hydrodynamic theories \citep{Mysak2006,V2006The,packham1968}.

We have experimentally investigated internal coastal Kelvin waves in a two-layer fluid system on a rotating table and focused on slow waves with frequencies lower than the table rotating frequency. Our experiments have shown that these waves, localized near the tank boundary, propagate in the same direction as the table rotation and decay exponentially along the propagation path. Wave dispersion relation from the inviscid theory has been experimentally verified with low-frequency waves when their propagations is nearly unaffected by dissipation. Protruding objects, including a half-cylinder and a perpendicular plate, are used as obstacles to modulate the wave propagation. Our experiments show that low-frequency waves can robustly propagate along the complex boundary (containing protruding objects) without being scattered and that the protruding objects do not cause additional wave dissipation; similar observations have been made in other topologically protected boundary states \citep{Nash2015,soni2019,Delplace2017,tauber2019,delplace2020}. However, as stressed in \citep{Tauber2020}, the topological origin of coastal Kelvin wave remains to be theoretically established with the corresponding topological invariant. Our experimental results and previous theoretical analysis suggest that fluid dynamical systems may provide a fertile platform to further extend the applications of topological methods and ideas.

\appendix

\begin{acknowledgments}
We acknowledge financial support from National
Natural Science Foundation of China (Grants No. 11774222, No. 12074243, and No. 11422427) and from the Program for Professor of Special Appointment at Shanghai Institutions of Higher Learning (Grant No. GZ2016004). The experimental studies at Tongji University were supported by the National Natural  Science Foundation of China (Grant No. 11772235) and a NSFC/RGC Joint Research (Grant No. 11561161004). We thank Mingji Huang and Siyuan Yang for useful discussions.
\end{acknowledgments}

\end{document}